\documentclass[aps,amssymb,amsmath,
twocolumn,showpacs,superscriptaddress,
groupedaddress]{revtex4-1}

\usepackage{graphicx}
\usepackage{dcolumn}
\usepackage{bm}
\usepackage[table,xcdraw]{xcolor}
\usepackage{nth}
\usepackage[ruled,vlined]{algorithm2e}
\usepackage{framed}
\usepackage{listings}
\usepackage{tikz}
\usetikzlibrary{shapes.geometric, arrows}
\tikzstyle{startstop} = [rectangle, rounded corners, minimum width=3cm, minimum height=1cm,text centered, draw=black, fill=white!30]
\tikzstyle{io} = [trapezium, trapezium left angle=70, trapezium right angle=110, minimum width=3cm, minimum height=1cm, text centered, draw=black, fill=white!30]
\tikzstyle{process} = [rectangle, minimum width=3cm, minimum height=1cm, text centered, draw=black, fill=white!30]
\tikzstyle{decision} = [diamond, minimum width=3cm, minimum height=1cm, text centered, draw=black, fill=white!30]
\tikzstyle{arrow} = [thick,->,>=stealth]
\begin{document}
\title{An efficient approximation for accelerating convergence of the numerical power series. Results for the $1D-$Schr{\"o}dinger's equation}
\author{A. Ba{\u g}c{\i}}
\email{abagci@pau.edu.tr}
\affiliation{Computational Physics Laboratory, Department of Physics, Faculty of Arts and Sciences, Pamukkale University, {\c C}amlaraltı, K{\i}n{\i}kl{\i} Campus, Denizli, Turkey}
\author{Z. G{\"u}ne{\c s}}
\affiliation{Department of Chemical Engineering, Faculty of Engineering, Pamukkale University, {\c C}amlaraltı, K{\i}n{\i}kl{\i} Campus, Denizli, Turkey}
\begin{abstract}
The numerical matrix Numerov algorithm is used to solve the stationary Schr{\"o}dinger equation for central Coulomb potentials. An efficient approximation for accelerating the convergence is proposed. The Numerov method is error$-$prone if the magnitude of grid$-$size is not chosen properly. A number of rules so far, have been devised. The effectiveness of these rules decrease for more complicated equations. Efficiency of the technique used for accelerating the convergence is tested by allowing the grid$-$sizes to have variationally optimum values. The method presented in this study eliminates the increased margin of error while calculating the excited states. The results obtained for energy eigenvalues are compared with the literature. It is observed that, once the values of grid$-$sizes for hydrogen energy eigenvalues are obtained, they can simply be determined for the hydrogen iso$-$electronic series as, $h_{\varepsilon}(Z)=h_{\varepsilon}(1)/Z$.
\begin{description}
\item[Keywords]
Numerov method, Screened Columb potential, Schr{\"o}dinger equation
\item[PACS numbers]
... .
\end{description}
\end{abstract}
\maketitle

\section{Introduction} \label{introduction}
The wave$-$function that holds all the required information of a system is obtained from analytical or numerical solution of the quantum mechanical wave$-$equation such as Schr{\"o}dinger or Dirac equation. Characteristics of the potential determines an exact analytical solution for the wave$-$equation whether available or not. If the potential function $V (r, r_{0})$ is separable depending on its variables $r,r_{0}$ into two factors \cite{1_Nichols_1965} may be reduced to soluble form. Such simplicity however, appears only for a few idealized systems. Complexity increases and the potential function may not be separable for more realistic representations. In this case, an analytical solution based on the approximation methods is required. Among these approximations, the perturbation, variation and Wentzel-Kramers-Brillouin (WKB) methods have generally been used in the literature \cite{2_Ishikawa_2002, 3_Landau_1977}. Although these analytical approaches are preferred for use in atomic and molecular calculations, they are limited in terms of applications \cite{2_Ishikawa_2002}. A solution free from symmetry properties of a potential inevitably leads to numerical approaches \cite{4_Dijk_2016}.\\
Numerical integration formula for integrating the time$-$independent Schr{\"o}dinger equation \cite{5_Uria_1996}
\begin{align}\label{eq1}
-\frac{\hbar}{2m}\psi''\left(x\right)+V\left(x\right)\psi\left(x\right)\approx E\psi\left(x\right)
\hspace{2mm} x \in \left[a,b\right]
\end{align}
which was suggested in the early days of quantum mechanics by Numerov \cite{6_Numerov_1924, 7_Numerov_1927}. The Numerov's method takes advantage of specific structure of the Eq. (\ref{eq1}). It involves no first derivative and it is linear in $\psi$. The Eq. (\ref{eq1}) may be transformed to dimensionless form as,
\begin{align}\label{eq2}
\psi''\left(x\right)-f\left(x\right)\psi\left(x\right)=0,
\end{align}
where, $f\left(x\right)=-2m\left(E-V\left(x\right)\right)/\hbar^2$.\\
The solution $\psi\left(x\right)$ at the nodes of $N$ uniform grid points, namely, $x_{0}, x_{1}, x_{2},...,x_{N}$ with $h_{\varepsilon}=x_{i+1}-x_{i}$, where $h_{\varepsilon}$ grid$-$size for $\varepsilon^{th}$ energy state. The Numerov's algorithm includes higher order terms than the well$-$known finite$-$difference with an error term that is of order $\mathcal{O}\left(h^2\right)$. It is derived by considering conjointly the forward, backward Taylor expansion of $\psi\left(x\right)$ up to order $\mathcal{O}\left(h^6\right)$ \cite{8_Graen_2016}. The following relation is finally obtained $\left[\psi_{i} \equiv \psi\left(x_{i}\right), f_{i}\equiv f_{i}\left(x\right)\right]$,
\begin{multline}\label{eq3}
\psi_{i+1}-2\psi_{i}+\psi_{i-1}=\\
-\frac{h^2}{12}\left(f_{i+1}\psi_{i+1}+10f_{i}\psi_{i}+f_{i-1}\psi_{i-1}\right) + \mathcal{O}\left(h^6\right).
\end{multline}
The Eq. (\ref{eq3}) is a numerical one$-$dimensional pattern \cite{9_Chow_1972}. The wavefunction $\psi\left(x\right)$ is calculated iteratively \cite{10_Gonzales_1997}. An initial guess for $E$ is required. The interval $\left[a,b\right]$ is finite and implicit the Dirichlet boundary conditions at $\psi_{0}=\psi\left(a\right)=0$, $\psi_{N}=\psi\left(b\right)=0$ \cite{8_Graen_2016}. Richardson extrapolation \cite{5_Uria_1996} or Cooley's correction formula \cite{11_Cooley_1961,12_Guest_1974, 13_Du_1990, 14_Izaac_2018} which are based on perturbation theory, may be used to improve the approximation for the energy with each iteration. \\
Several other variants in order to increase both stability and accuracy while calculating the resonance problems and high$-$lying bound states were proposed. Generalization of the algorithm to an error of arbitrary order \cite{15_Hajj_1974, 16_Fack_1987, 17_Berghe_1989, 18_Ixaru_1980, 19_Ixaru_1985, 20_Allison_1991, 21_Killingbeck_1999, 22_Wang_2004, 23_Yang_2017, 24_Obaidat_2021, 25_Medvedeva_2021} and its extension for solution of differential equations with more than one$-$dimension \cite{8_Graen_2016,26_Kobeisse_1974, 27_Hajj_1975, 28_Hajj_1982, 29_Hajj_1985, 30_Eckert_1989, 31_Avdelas_2000, 32_Kalogiratou_2005, 33_Kuenzer_2019} form the main framework of the studies. A Numerov$-$type exponentially fitted method was suggested \cite{18_Ixaru_1980, 19_Ixaru_1985, 20_Allison_1991, 32_Kalogiratou_2005, 34_Raptis_1978, 35_Berghe_1990, 36_Berghe_1990, 37_Simos_1996, 38_Simos_1998, 39_Simos_1999, 40_Konguetsof_2003, 41_Aguiar_2005, 42_Berghe_2007, 43_Tsitouras_2018}, accordingly. Here, the coefficients in the Eq. (\ref{eq3}) are replaced by some arbitrary parameters. It is assumed that the solution is a linear combination of the exponential functions. The parameters are then obtained from the resulting system of differential equations. Another method derived from the Eq. (\ref{eq3}) is so called re-normalized Numerov method. It was obtained by making two transformations \cite{44_Johnson_1977, 45_Johnson_1978, 46_Leroy_1985, 47_Karman_2014, 48_Zhao_2016}. The first transformation is used to decrease the number of steps required for calculation and the second one is used to replace the three$-$point recurrence relation with two$-$point relations by defining a ratio. If the iteration is stopped at any point, the last two elements of the ratio are immediately available. They are used to calculate the wave$-$function $\left(\psi \right)$ around that point within a normalized factor \cite{44_Johnson_1977}. Non$-$uniform grid$-$size $h_{i}$, $h_{i+1}=x_{i+1}-x_{i}$ was also considered to be use \cite{49_Kobeissi_1988, 50_Bieniasz_2004, 51_Aguiar_2005, 52_Ramos_2005, 53_Speciale_2020, 54_Brunetti_2021}. It is derived due to the need to solve the transport model for semiconductor devices where, the standard Numerov procedure with uniform grid$-$size is not applicable \cite{53_Speciale_2020, 54_Brunetti_2021}. The range of application of the Numerov's algorithm is indubitably not limited to the aforementioned examples \cite{47_Karman_2014, 48_Zhao_2016, 50_Bieniasz_2004, 53_Speciale_2020, 54_Brunetti_2021}. It is used to determine presence of the energetics for short intramolecular O$-$H$\cdots$O in enzyme and photo centers by nuclear motion of the involved hydrogen atom \cite{8_Graen_2016}, to solve the Schr{\"o}dinger equation with complex potential boundaries for open multi$-$layer heterojunction systems \cite{55_Lin_2018} and to study energy spectra of mesons and hadrons \cite{56_Ali_2020}. Two$-$dimensional quantum dot eigen$-$functions for a much larger spectrum of external harmonic frequencies are calculated as well \cite{57_Caruso_2019}.

As an alternative to the iterative procedure the Numerov's algorithm can be transformed to a matrix form. This matrix method was suggested in \cite{5_Uria_1996} and was revisited in \cite{58_Mohandas_2012}. It leads to renewed interest in solution of the one$-$dimensional Schr{\"o}dinger equation. Convergence properties for the Numerov's algorithm has been recently investigated for various forms of central potentials \cite{59_Xie_2021} where, the relativistic effects have also been taken into account. In this matrix method, the problem is reduced to the following form of the algebraic generalized$-$eigenvalue equation,
\begin{align}\label{eq4}
Ax=\lambda B x.
\end{align}
Computation of the energy levels and the wave$-$functions in the Eq. (\ref{eq1}) is now reduced to the computation of eigenvalues and eigen$-$functions of triangular symmetric matrices $A$, $B$. This leads to the ground and the excited states of the Eq. (\ref{eq1}) to be calculated simultaneously. No initial guess is required \cite{5_Uria_1996}. Efficient computation of the Eq. (\ref{eq4}) for matrices with large numbers of non$-$zero elements has attracted considerable research interest because it is encountered in many applications in physics, chemistry and engineering. Gaussian elimination, $LU-$factorization, inverse iteration methods are the ones were studied in detail \cite{60_Stewart_1973}.

The proposed approach in this work can be considered within the scope of matrix Numerov method. The number of grid points is reduced while the accuracy is sufficiently increased. This is achieved by manipulating the potential in the Schr{\"o}dinger equation. Due to singularity of the Coulomb potential, hydrogen-like eigen$-$functions have non$-$continuous derivatives at $r=0$ \cite{61_Fattal_1996}. Correctly representation of the radial wave$-$function require use of tremendous grid points such that, one has to consider how to compute Eq. (\ref{eq4}) effectively. The solution of the Schr{\"o}dinger equation for the central Coulomb potential is therefore examined as a sample. The error term is embedded in a certain weight to the matrix that represents the effective potential. Such weight leads to approximately determine the grid$-$size. Variational stability is tested via the optimization procedure. The results are presented for hydrogen$-$like atoms.
%%%%%%%%%%%%%%%%%%%%%%%%%%%%%%%%%%%%%%%%%%%%%%%%%%%%%%%%%%%%%%%%%%%%%%%
%%%%%%%%%%%%%%%%%%%%%%%%%%%%%%%%%%%%%%%%%%%%%%%%%%%%%%%%%%%%%%%%%%%%%%%
%%%%%%%%%%%%%%%%%%%%%%%%%%%%%%%%%%%%%%%%%%%%%%%%%%%%%%%%%%%%%%%%%%%%%%%
\begin{table*}
\caption{\label{tab:h} Variationally optimum values of grid$-$sized to be used in computation of electronic energy spectrum of hydrogen atom. The calculations are performed in atomic units (a. u.) for the pure Coulomb potential where, $\nu=0$ and $\mu\left(N,0,0\right)=0$.}
\begin{ruledtabular}
\begin{tabular}{ccccccc}
$N$ & $h_{1s}$ & $h_{2s}$ & $h_{3s}$ & $h_{4s}$ & $h_{5s}$ & $h_{10s}$
\\
\hline
$500$ & $0.016764$ & $0.041090$ & $0.073143$ & $0.112906$ & $0.160429$ & $0.518061$
\\
$1000$ & $0.009129$ & $0.022209$ & $0.039261$ & $0.060262$ & $0.085171$ & $0.268797$
\\
$1500$ & $0.006378$ & $0.015458$ & $0.027225$ & $0.041659$ & $0.058729$ & $0.183493$
\\
$2000$ & $0.004941$ & $0.011920$ & $0.020973$ & $0.032028$ & $0.045083$ & $0.140004$
\\
$2500$ & $0.004048$ & $0.009749$ & $0.017134$ & $0.026108$ & $0.036703$ & $0.113499$
\\
$5000$ & $0.002171$ & $0.005199$ & $0.009070$ & $0.013798$ & $0.019333$ & $0.059092$
\\
$7500$ & $0.001498$ & $0.003590$ & $0.006244$ & $0.009485$ & $0.013268$ & $0.040302$
\\
$10000$ & $0.001160$ & $0.002755$ & $0.004798$ & $0.007261$ & $0.010146$ & $0.030702$
\\
$15000$ & $0.000804$ & $0.001902$ & $0.003286$ & $0.004979$ & $0.006947$ & $0.020917$
\end{tabular}

\end{ruledtabular}
\end{table*}

\begin{table*}
\caption{\label{tab:eh} Variationally minimum values for electronic energy spectrum of hydrogen atom with the Coulomb potential in atomic units (a. u.) where, $\nu=0$ and $\mu\left(N,0,0\right)=0$.}
\begin{ruledtabular}
\begin{tabular}{ccccccc}
$N$ & $E_{1s}\left(N,h_{1s},0\right)$ & $E_{2s}\left(N,h_{2s},0\right)$ & $E_{3s}\left(N,h_{3s},0\right)$ & $E_{4s}\left(N,h_{4s},0\right)$ & $E_{5s}\left(N,h_{5s},0\right)$ & $E_{10s}\left(N,h_{10s},0\right)$
\\
\hline
$500$ & \textbf{0.499}89 54423 & \textbf{0.1249}2 43874 & \textbf{0.055}48 76675 & \textbf{0.031}18 52044 & \textbf{0.0199}3 68695 & \textbf{0.0049}4 40578
\\
$1000$ & \textbf{0.4999}6 90692 & \textbf{0.1249}7 76755 & \textbf{0.0555}3 54185 & \textbf{0.0312}3 05773 & \textbf{0.0199}8 07819 & \textbf{0.0049}8 07719
\\
$1500$ & \textbf{0.4999}8 49515 & \textbf{0.1249}8 91750 & \textbf{0.0555}4 57959 & \textbf{0.0312}4 05672 & \textbf{0.01999} 06272 & \textbf{0.00499} 01987
\\
$2000$ & \textbf{0.49999} 10010 & \textbf{0.12499} 35460 & \textbf{0.0555}4 97440 & \textbf{0.0312}4 43816 & \textbf{0.01999} 44088 & \textbf{0.00499} 40255
\\
$2500$ & \textbf{0.49999} 39700 & \textbf{0.12499} 56864 & \textbf{0.05555} 16764 & \textbf{0.0312}4 62510 & \textbf{0.01999} 62672 & \textbf{0.00499} 59622
\\
$5000$ & \textbf{0.49999} 82742 & \textbf{0.12499} 87763 & \textbf{0.05555} 44612 & \textbf{0.0312}4 89454 & \textbf{0.01999} 89507 & \textbf{0.00499} 88427
\\
$7500$ & \textbf{0.49999 9}1736 & \textbf{0.12499 9}4173 & \textbf{0.05555 5}0364 & \textbf{0.03124 9}5010 & \textbf{0.01999 9}5043 & \textbf{0.00499 9}4518
\\
$10000$ & \textbf{0.49999 9}5110 & \textbf{0.12499 9}6564 & \textbf{0.05555 5}2504 & \textbf{0.03124 9}7072 & \textbf{0.01999 9}7095 & \textbf{0.00499 9}6789
\\
$15000$ & \textbf{0.49999 9}7669 & \textbf{0.12499 9}8372 & \textbf{0.05555 5}4115 & \textbf{0.03124 9}8622 & \textbf{0.01999 9}8636 & \textbf{0.00499 9}8498
\\
&
\begin{tabular}{cc}
\hspace{0.5mm} \textbf{0.49999 99}167\footnotemark[1]
\\
\textbf{0.49999} 0\footnotemark[2]
\end{tabular} 
&
\begin{tabular}{cc}
\hspace{0.5mm} \textbf{0.12499 99}896\footnotemark[1]
\\
\textbf{0.124999}\footnotemark[2]
\end{tabular}
&
\textbf{0.05555} 4\footnotemark[2]
& 
& 
& 
\end{tabular}

\end{ruledtabular}
\footnotetext[1]{Ref. \cite{69_Purevkhuu_2021} $\left(N \approx 25000 \right)$}
\footnotetext[2]{Ref. \cite{5_Uria_1996}}
\end{table*}
%%%%%%%%%%%%%%%%%%%%%%%%%%%%%%%%%%%%%%%%%%%%%%%%%%%%%%%%%%%%%%%%%%%%%%%
%%%%%%%%%%%%%%%%%%%%%%%%%%%%%%%%%%%%%%%%%%%%%%%%%%%%%%%%%%%%%%%%%%%%%%%
%%%%%%%%%%%%%%%%%%%%%%%%%%%%%%%%%%%%%%%%%%%%%%%%%%%%%%%%%%%%%%%%%%%%%%%
\begin{table*}
\caption{\label{tab:eih} Results of computation for electronic energy spectrum of hydrogen atom with screened Coulomb potential in atomic units (a. u.) where, $\nu \neq 0$ and $\mu\left(N,0,\nu\right) \neq 0$.}
\begin{ruledtabular}
\begin{tabular}{ccccccc}
$N$ & $E_{1s}\left(N,h_{1s},\nu\right)$ & $E_{2s}\left(N,h_{2s},\nu\right)$ & $E_{3s}\left(N,h_{3s},\nu\right)$ & $E_{4s}\left(N,h_{4s},0\right)$ & $E_{5s}\left(N,h_{5s},\nu\right)$ & $E_{10s}\left(N,h_{10s},\nu\right)$
\\
\hline
$500$ & \textbf{0.49999} 81387 & \textbf{0.12499} 35304 & \textbf{0.05555} 00986 & \textbf{0.0312}4 43978 & \textbf{0.01999} 43970 & \textbf{0.00499} 46365
\\
$1000$ & \textbf{0.49999 9}5666 & \textbf{0.12499} 81450 & \textbf{0.05555} 39025 & \textbf{0.0312}4 81011 & \textbf{0.01999} 78112 & \textbf{0.00499} 57720
\\
$1500$ & \textbf{0.49999 9}8415 & \textbf{0.12499 9}1616 & \textbf{0.05555} 48140 & \textbf{0.03124 9}1164 & \textbf{0.01999} 89347 & \textbf{0.00499 9}5057
\\
$2000$ & \textbf{0.49999 99}394 & \textbf{0.12499 9}5392 & \textbf{0.05555 5}1559 & \textbf{0.03124 9}5120 & \textbf{0.01999 9}3941 & \textbf{0.00499 9}6099
\\
$2500$ & \textbf{0.49999 99}699 & \textbf{0.12499 9}7087 & \textbf{0.05555 5}3086 & \textbf{0.03124 9}6943 & \textbf{0.01999 9}6129 & \textbf{0.00499 9}7098
\\
$5000$ & \textbf{0.50000 000}10 & \textbf{0.12499 99}336 & \textbf{0.05555 55}062 & \textbf{0.03124 99}361 & \textbf{0.01999 99}133 & \textbf{0.00499 99}207
\\
$7500$ & \textbf{0.49999 999}60 & \textbf{0.12499 99}684 & \textbf{0.05555 55}341 & \textbf{0.03124 99}728 & \textbf{0.01999 99}626 & \textbf{0.00499 99}651
\\
$10000$ & \textbf{0.50000 000}37 & \textbf{0.12499 99}866 & \textbf{0.05555 55}485 & \textbf{0.03124 99}899 & \textbf{0.01999 99}842 & \textbf{0.00499 99}865
\\
$15000$ & \textbf{0.50000 000}36 & \textbf{0.12499 999}59 & \textbf{0.05555 555}48 & \textbf{0.03124 999}81 & \textbf{0.01999 999}56 & \textbf{0.00499 999}76
\\
 & $i=3$ & $i=1$ & $i=2$ & $i=4$ & $i=6$ & $i=15$
\end{tabular}
\footnotetext{$i$ is the number of iterations}
\end{ruledtabular}
\end{table*}
%%%%%%%%%%%%%%%%%%%%%%%%%%%%%%%%%%%%%%%%%%%%%%%%%%%%%%%%%%%%%%%%%%%%%%%
%%%%%%%%%%%%%%%%%%%%%%%%%%%%%%%%%%%%%%%%%%%%%%%%%%%%%%%%%%%%%%%%%%%%%%%
%%%%%%%%%%%%%%%%%%%%%%%%%%%%%%%%%%%%%%%%%%%%%%%%%%%%%%%%%%%%%%%%%%%%%%%
\section{Modified Matrix Numerov Representation for The Hamiltonian} \label{mmn}
Taking into account the Eqs. (\ref{eq1},\ref{eq2}) the exact second derivative for $\psi\left(x\right)$ is given as,
\begin{align}\label{eq5}
\psi''\left(x\right)=-\frac{2m}{\hbar^2}\left[E-V\left(x\right)\right]\psi\left(x\right).
\end{align}
The Eq. (\ref{eq5}) from the perspective of numerical integration, may be represented by the following statement;
\begin{align*}
Exact=Numerical+\mathcal{O}\left(h^6\right).
\end{align*}
By inserting the Eq. (\ref{eq5}) in the above statement and assuming that the error term is of order $\mathcal{O}\left(h^6\right)$ we have the following property,
\begin{align}\label{eq6}
\hat{\beta}\left(h^6\right)\psi\left(x\right)
=\mathcal{O}\left(h^6\right)\psi\left(x\right),
\end{align}
\begin{align}\label{eq7}
Numerical=-\frac{2m}{\hbar^2}\left[E-V\left(x\right) \pm \beta\left(h^6\right)\right]\psi\left(x\right).
\end{align}
If the potential term defined as,
\begin{align}\label{eq8}
V'\left(x\right)=V\left(x\right) \pm \beta\left(h^6\right),
\end{align}
the Eq. (\ref{eq7}) becomes analogous to Eq. (\ref{eq2}). It is unlikely to find a definition for $\beta\left(h^6\right)$. Instead the potential $V\left(x\right)$ may be replaced by the one involving a screening parameter $\mu$ (more precisely a function). Thus,
\begin{align}\label{eq9}
V'\left(x\right)=V_{\mu}\left(x\right).
\end{align}
In general we can say that the function $\mu$ is a function of number of grid$-$points. It must be a function such that it satisfies the conditions given for the screened potential as below,
\begin{align}\label{eq10}
\lim_{\mu\rightarrow 0}V'_{\mu}\left(x\right)=V\left(x\right),
\end{align}
\begin{align}\label{eq11}
\lim_{N\rightarrow \infty}V'_{\mu}\left(x\right)=V\left(x\right).
\end{align}
%The illustration for the Eqs. (\ref{eq11},\ref{eq12}) is also given in the \textcolor{red}%{Figure 1}.
We have derived two variants of $\mu$ in order to solve the radial stationary Schr{\"o}dinger equation of an electron moving through the Coulomb potential.
\section{A method to determine the function $\mu\left(N\right)$}\label{mu}
Depending on the number of grid$-$points $N$ and angular momentum quantum number $l$, the two versions of the $\mu$ function in this study are defined as,
\begin{align}\label{eq12}
\mu_{1}\left(N,l,\nu_{1}\right)=Exp\left[{\frac{1}{N^{\nu_{1}(l+1)}}}\right]-1,
\end{align}
\begin{align}\label{eq13}
\mu_{2}\left(N,l,\nu_{2}\right)=Erf\left[{\frac{1}{N^{\nu_{2}(l+1)}}}\right],
\end{align}
where, $\left\lbrace \nu_{1}, \nu_{2} \right\rbrace \in \mathcal{R}^{+}$, $1 \leq \left\lbrace \nu_{1}, \nu_{2} \right\rbrace < \infty$.\\
Minimum and maximum values to be found for energy eigenvalues correspond to the $\nu=1$ and $\nu \rightarrow \infty$ ($\mu\rightarrow 0$, represents the pure Coulomb potential), respectively. According to the value of $\nu$ the interval for numerically obtained energy eigenvalues are determined as,
\begin{align*}
E_{1} \leq E_{\nu} \leq E_{\infty}=E_{Coulomb}.
\end{align*}
The Eqs. (\ref{eq12}, \ref{eq13}) for any value of angular momentum quantum number $l$ satisfy the following property,
\begin{align}\label{eq14}
\lim_{N\rightarrow \infty}\mu\left(N,l,\nu\right)=0.
\end{align}
The Eq. (\ref{eq4}) is solved iteratively since it is now depends to the values of $\nu$. The number of iterations are sensitive to the choice of initial values of $\nu$. Large number of grid$-$points approximate both values of the $\mu$ and the grid$-$size $h$ to zero. A simple approach as $\mu_{1}\left(N,l,\nu_{1}\right)=h_{1s}^2$, $\mu_{2}\left(N,l,\nu_{2}\right)=h_{1s}^2$ or visa versa (see the Table \ref{tab:h}) that promise better initial values for $\nu$ (or if $\nu$ is known for $h$) is used, accordingly. Alternatively, one can chose a large value of $\nu$ as an initial input. 
Two upper limit of summations (number of grid$-$points) are selected. The calculations are performed for these two upper limit of summations using the $\nu_{1}$. If the energy eigenvalues are increasing, then the $\nu_{2}$ is used. The iteration is terminated. Else, $\nu_{1}$ is used in the Eq. (\ref{eq13}) to derive a new $\nu_{1}$. This new value of $\nu_{1}$ is used in the Eq. (\ref{eq12}). The calculations are performed until the energy eigenvalues start to increasing. An algorithm for the above procedure may be summarized as,\\ \\
\textbf{Step 1.} Choose two upper limit of summation $\left(N_{1}, N_{2}\right)$.\\
\textbf{Step 2.} Choose a grid$-$size for each upper limit of summation or use the ones given in the Table \ref{tab:h}. \\
\textbf{Step 3.} Obtain initial values of $\nu_{1}$ and $\nu_{2}$ for each $N$ by solving the equations $\mu_{1}\left(N,l,\nu_{1} \right)=h^2$, $\mu_{2}\left(N,l,\nu_{2} \right)=h^2$. $i=0$ (iteration number).\\
\textbf{Step 4.} Calculate the energy eigenvalues $E_{N_1}\left(h_{1},\nu_{1}\right)$, $E_{N_2}\left(h_{2},\nu_{1}\right)$. \\
\textbf{Step 5.} If $E_{N_1}\left(h_{1},\nu_{1}\right)<E_{N_2}\left(h_{2},\nu_{1}\right)$ and $i=0$\\
Print $E_{N_2}\left(h_{2},\nu_{2}\right)$. Break.\\
\textbf{Step 6.} Else if $E_{N_1}\left(h_{1},\nu_{1}\right)>E_{N_2}\left(h_{2},\nu_{1}\right)$\\
Insert the $\nu_{1}$ in the Eq. (\ref{eq13}). Obtain new values for $\mu_{2}$. Solve the equation $\mu_{1}\left(N,l,\nu_{1} \right)=\mu_{2}$. This derives a new value for $\nu_{1}$. $i=i+1$. Go to step 4.\\
\textbf{Step 7.} Write,\\
\begin{align*}
\frac{\nu^{2,i-1}_{1} E^{i-1}_{N_{2}}+\nu^{2,i}_{1} E^{i}_{N_{2}}}{\nu^{2,i-1}_{1}+\nu^{2,i}_{1}}.
\end{align*}
Note that $E^{i}_{N} \equiv E^{i}_{N}\left(h,\nu\right) \equiv E_{nl}\left[h,\mu\left(N,l,\nu\right) \right]$ of $i^{th}$ iteration. $n=1,2,3,...$, $l=n-1$, represent the corresponding energy state. $\nu_{1}^{p i}$, $\nu_{2}^{p i}$ represent $i^{th}$ values of $\nu_{1}$ and $\nu_{2}$ for grid$-$points $N_{1}, N_{2}$ with $p$, $p=1$ or $p=2$, respectively. $h_{1}$ and $h_{2}$ are the grid$-$sizes for the corresponding energy eigenvalues in $N_{1}$, $N_{2}$. The convergence properties and the algorithm of the procedure with an explicit application are given in the Figures \ref{fig:E4s}, \ref{fig:chart}.
\section{Example: Hydrogen$-$like atoms} \label{hlike}
The radial Schr{\"o}dinger equation for an electron in a spherically symmetric potential has the following form \cite{3_Landau_1977},
\begin{align}\label{eq15}
\hat{H}_{r}\Psi\left(r\right)=E_{r}\Psi\left(r\right),
\end{align}
where,
\begin{align}\label{eq16}
\hat{H}_{r}=-\frac{1}{2}\Bigg(\frac{\partial^2}{\partial r^2}-\frac{l(l+1)}{r^2}\Bigg)+\frac{1}{r},
\end{align}
is the one$-$electron Hamilton operator [in atomic units (a. u.); $\hbar=1$,$m=1$and $e^2/4\pi\epsilon_{0}=1$].
Note that, the Eq. (\ref{eq16}) depends upon $l$ yet, the exact solution of the Eq. (\ref{eq15}) with $\hat{H}_{r}$ for hydrogen atom is $n^2-$\textit{fold} degenerate. It is referred to as \textit{accidental degeneracy}. The discussion on the origin of such degeneracy for hydrogen atom is treated rigorously in \cite{62_Deshmukh_2015}. The results obtained from numerical solution of the Schr{\"o}dinger equation using the standard Numerov's method for $\hat{H}_{r}$ operator do not posses $n^2-$\textit{fold} degeneracy (see the following section for more detail). The Numerov's method with a modified effective potential on the other hand, reveals the \textit{accidental degeneracy}.

In order to apply the Numerov method with a modified effective potential, the Eq. (\ref{eq15}) is transformed into a linear finite$-$difference equation. The effective potential is defined as,
\begin{align}\label{eq17}
V' \equiv V'_{eff}\left(r\right)=\frac{1}{2}\frac{\left[l(l+1)\right]}{r^2}+\frac{1}{r^{1-\mu}}.
\end{align}
The standard Numerov algorithm given in the Eq. (\ref{eq3}) is used in following. The resulting equation can be rearranged into the following form $\left(\Psi \equiv \Psi\left(r\right)\right)$ \citep{5_Uria_1996, 58_Mohandas_2012},
\begin{multline}\label{eq18}
-\frac{1}{2}\frac{\left(\Psi_{i-1}-2\Psi_{i}+\Psi_{i+1}\right)}{h^2}\\
+\frac{\left(V'_{i-1}\Psi_{i-1}+10V'_{i}\Psi_{i}+V'_{i+1}\Psi_{i+1}\right)}{12}\\
=E\frac{\left(\Psi_{i-1}+10\Psi_{i}+\Psi_{i+1}\right)}{12}.
\end{multline}
$N$ equations of the linear system given above in the matrix form are written as,
\begin{align}\label{eq19}
-\frac{1}{2}A\Psi+BV' \Psi=EB\Psi,
\end{align}
where, $\Psi$ as the column vector $\left(...\Psi_{i-1},\Psi_{i},\Psi_{i+1}...\right)$, $A=\left(\mathbb{I}_{-1}-2\mathbb{I}_{0}+\mathbb{I}_{1} \right)/h^2$, $B=\left(\mathbb{I}_{-1}-2\mathbb{I}_{0}+\mathbb{I}_{1} \right)/12$, $V'=diag\left(...V'_{i-1},V'_{i},V'_{i+1}...\right)$, $\mathbb{I}_{p}$ is a matrix of $1s$ along the $pth$ diagonal and zeros elsewhere. The boundary conditions $\Psi\left(0\right)=\Psi\left(R\right)=0$ for some large $R$ are implemented by taking $N\times N$ sub$-$matrices of $A$ and $B$. The optimized values of $h$ is obtained by minimizing the energy eigenvalues. Then, the solution for $\Psi$ is found at the interval $\left[0,R \right]$, $R=N \times h$.
\section{Results and Discussions} \label{rd}
The modified Numerov algorithm proposed in the present paper is based on the potential function has a screening parameter, $\mu$. Such a potential is first suggested to calculate the matrix elements of molecular properties such as quadrupole coupling tensor, electron$-$photon hyperfine interaction, chemical shift and spin$-$orbit coupling  \cite{63_Silverstone_1971}. Later on, an exponential function was embedded into this potential function \cite{64_Guseinov_1988}. It was referred to as non$-$central interaction potential \cite{65_Guseinov_2002}. It is defined as,
\begin{align}\label{eq20}
f_{\mu\kappa\sigma}\left(\zeta,\vec{r}\right)
=r^{\mu-1}e^{-\zeta r}\left(\frac{4\pi}{2\kappa+1}\right)^{1/2}
S_{\kappa\sigma}\left(\theta,\varphi\right),
\end{align}
where $\mu>=-\left(\kappa-1\right)$ and $\kappa>=0$, $S_{\kappa\sigma}\left(\theta,\varphi\right)$ are the real or complex spherical harmonics. The case $\kappa=\sigma=0$ corresponds to the screened central potential. It was also considered that $\mu$ can take non$-$integer values at the range of $-1\leq \mu \leq 0$. In this case the potential referred to as correlated interaction potential in which it was claimed that the electron correlation effects are directly involved \cite{66_Guseinov_2008}. This type of potential was used to improve Hartree$-$Fock self$-$consistent field calculations \cite{67_Guseinov_2008}. It is on the other hand, obvious that the real values of $\mu$ may cause variational instability.

Based on the perspective of numerical integration, the aforementioned method in this study regards $\mu$ as a function enables the exact value for the second order derivative of $\psi\left(x\right)$ to be acquired at a finite number of grid$-$points. The function $\mu$ is chosen depending on the conditions given in the Eq. (\ref{eq10}, \ref{eq11}) and the Eq. (\ref{eq14}). The values of $\mu$ functions for certain number of grid$-$point are determined through the Eqs. (\ref{eq12}, \ref{eq13}). The Eqs. (\ref{eq12}, \ref{eq13}) depend on the angular momentum quantum number $l$ and their values significantly decrease by increasing the $l$.

\begin{figure*}
\caption{\label{fig:chart} The algorithm of the procedure given in the Section \ref{mu} with an example for ground state energy calculation of hydrogen atom. $N_{1}, N_{2}$ are the grid$-$points. $h_{1}, h_{2}$ are the grid$-$sizes for $N_{1}$, $N_{2}$. $i$ represents the number of iteration. $\nu_{1}^{p i}$, $\nu_{2}^{p i}$ represent $i^{th}$ of $\nu_{1}$ and $\nu_{2}$ for grid$-$points $N_{1}, N_{2}$ with $p$, $p=1$ or $p=2$, respectively.}
\centering
\begin{tikzpicture}[node distance=2cm]
\node (start) [startstop] {Start};
\node (in1) [io, below of=start,yshift=-0.3cm]{
\begin{tabular}{c}
$N_1=7500$ \\
$N_2=10000$\\
$h_1=0.001498$ \\
$h_2=0.001160$ \\
(Table \ref{tab:h})\\
$i=0$ (iteration number)
\end{tabular}};
\node (proc1) [process, below of=in1,yshift=-0.3cm] {
\begin{tabular}{c}
$\nu_{1}^{10}=1.457709$ \\
$\nu_{1}^{20}=1.467792$ 
\end{tabular}};
\node (proc2) [process, below of=proc1,yshift=-2.0cm] {
\begin{tabular}{c}
$0^{th}$ iteration, $i=0$ \\
$E^{0}_{N_1}\left(h, \nu_{1}^{10}\right)=-0.49999\hspace{1mm}97807$ \\
$E^{0}_{N_2}\left(h, \nu_{1}^{20}\right)=-0.49999\hspace{1mm}98747$ \\
\\
$1^{th}$ iteration, $i=1$ \\
$E^{1}_{N_1}\left(h, \nu_{1}^{11}\right)=-0.49999\hspace{1mm}98586$ \\
$E^{1}_{N_2}\left(h, \nu_{1}^{21}\right)=-0.49999\hspace{1mm}99214$ \\
\\
$2^{th}$ iteration, $i=2$ \\
$E^{2}_{N_1}\left(h, \nu_{1}^{12}\right)=-0.49999\hspace{1mm}99465$ \\
$E^{2}_{N_2}\left(h, \nu_{1}^{22}\right)=-0.49999\hspace{1mm}99741$ \\
\\
$3^{th}$ iteration, $i=3$ \\
$E^{3}_{N_1}\left(h, \nu_{1}^{13}\right)=-0.50000\hspace{1mm}00458$ \\
$E^{3}_{N_2}\left(h, \nu_{1}^{23}\right)=-0.50000\hspace{1mm}00336$ \\
\end{tabular}};
\node (dec1) [decision, below of=proc2, yshift=-3.1cm] {
\begin{tabular}{c}
$E^{i}_{N_1} < E^{i}_{N_2}$\\
$i=0$
\end{tabular} };
\node (proc3) [process, right of=dec1, xshift=2.3cm] {
\begin{tabular}{c}
$E^{0}_{N_2}\left(h, \nu_{2}^{20}\right)$
\end{tabular}};
\node (dec2) [decision, left of=dec1, xshift=-1.8cm] {$E^{i}_{N_1} > E^{i}_{N_2}$};
\node (proc4) [process, left of=dec2,xshift=-1.8cm] {
\begin{tabular}{c}
$i=i+1$\\
$1^{th}$ iteration, $i=1$\\
$\nu_{1}^{11}=1.444173$ \\
$\nu_{1}^{21}=1.454678$
\\ \\
$2^{th}$ iteration, $i=2$ \\
$\nu_{1}^{12}=1.430636$ \\
$\nu_{1}^{22}=1.441564$
\\ \\
$3^{th}$ iteration, $i=3$ \\
$\nu_{1}^{13}=1.417100$ \\
$\nu_{1}^{23}=1.428451$
\end{tabular}};
\node (out1) [io, below of=dec1,yshift=-0.3cm]{\begin{tabular}{c}
Print\\
$E=-0.50000\hspace{1mm}00037$
\end{tabular}};
\node (out2) [io, below of=proc3,yshift=-0.3cm]{\begin{tabular}{c}
Print \\
$E^{0}_{N_{2}}\left(h, \nu_{2}^{20} \right)$
\end{tabular}};
\draw [arrow] (start) -- (in1);
\draw [arrow] (in1) -- (proc1);
\draw [arrow] (proc1) -- (proc2);
\draw [arrow] (dec1) -- node[anchor=south] {yes} (proc3);
\draw [arrow] (dec1) -- node[anchor=south] {no} (dec2);
\draw [arrow] (proc2) -- (dec1);
\draw [arrow] (dec2) -- node[anchor=south] {yes} (proc4);
\draw [arrow] (dec2) |- node[anchor=north] {no} (out1);
\draw [arrow] (proc4) |- (proc2);
\draw [arrow] (proc3) -- (out2);
\node (stop) [startstop, below of=out1] {Stop};
\draw [arrow] (out2) |- (stop);
\draw [arrow] (out1) -- (stop);
\end{tikzpicture}
\end{figure*}
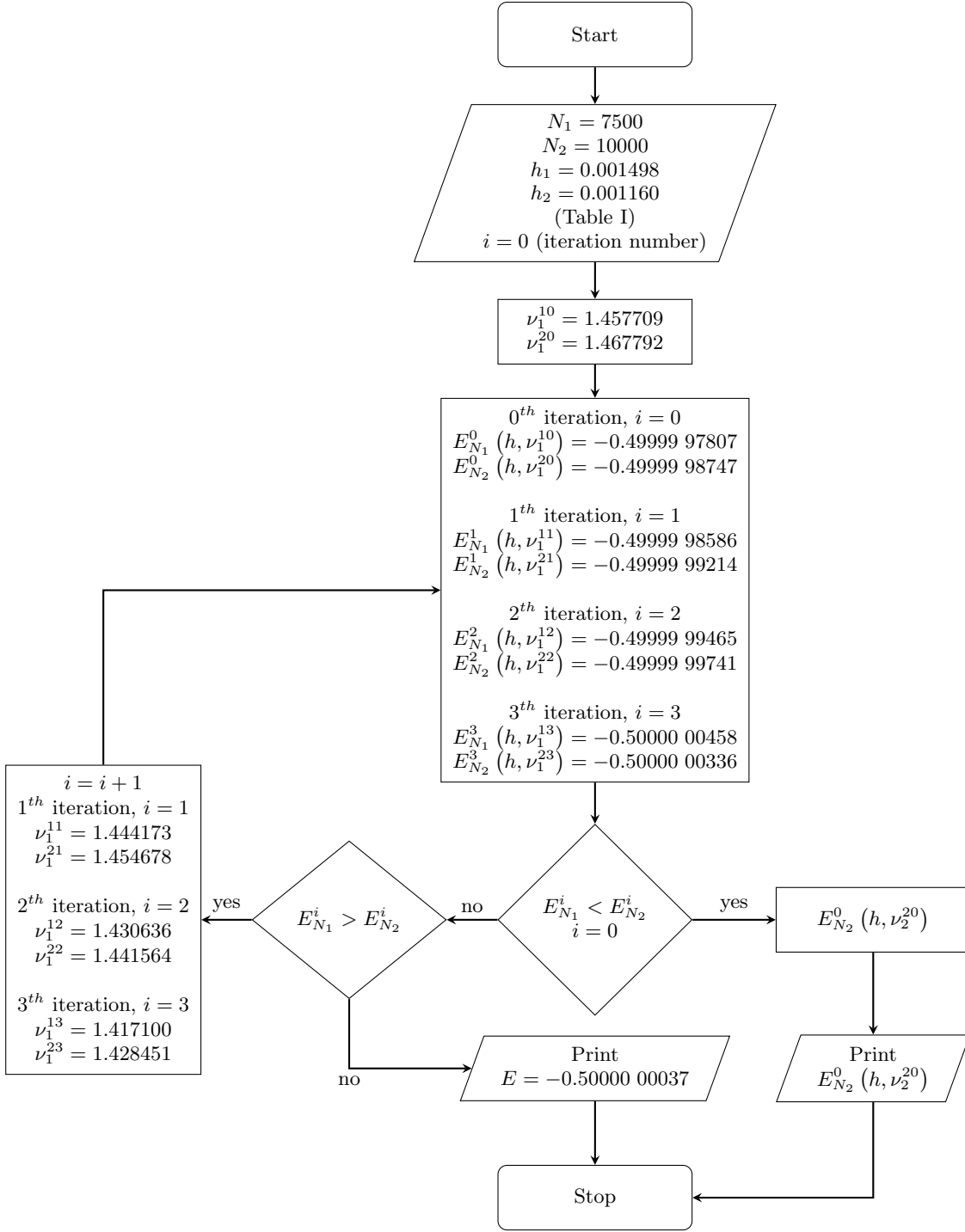

The energy eigenvalues are usually calculated by requiring the eigen$-$functions to satisfy certain boundary conditions \cite{68_Chin_2019}. The method presented in this study avoids dependence of energy eigenvalues to the eigen$-$functions. It suggests first to calculate the eigenvalues. The corresponding eigen$-$functions are then calculated easily. In this case all the energy eigenvalues should be tried up to a certain precision. This approximation was used previously in \cite{68_Chin_2019}. An infinite potential barrier at some radius $C$ added to a potential. The correct eigenvalue is then obtained in the limit $C\rightarrow \infty$. It is referred to as \textit{hardwall method}. The number of significant digits provided by this method is limited to a few. In this work, an explicit function that satisfy the conditions given in the Eqs. (\ref{eq10}, \ref{eq11}, \ref{eq14}) is alternatively added to the potential. The presented algorithm provides arbitrary number of significant digit because the range of values of $\mu$ indicates where the exact energy eigenvalue is (see Figure \ref{fig:chart}). This yields the variational instability to be eliminated as well. The parameter $\mu$ in the non$-$central interaction potential free from restriction and the energy eigenvalues are unbounded from below unless the Eqs. (\ref{eq10}, \ref{eq11}, \ref{eq14}) are taken into account.
\begin{figure}[ht]
\includegraphics[width=0.47\textwidth,height=0.30\textheight]{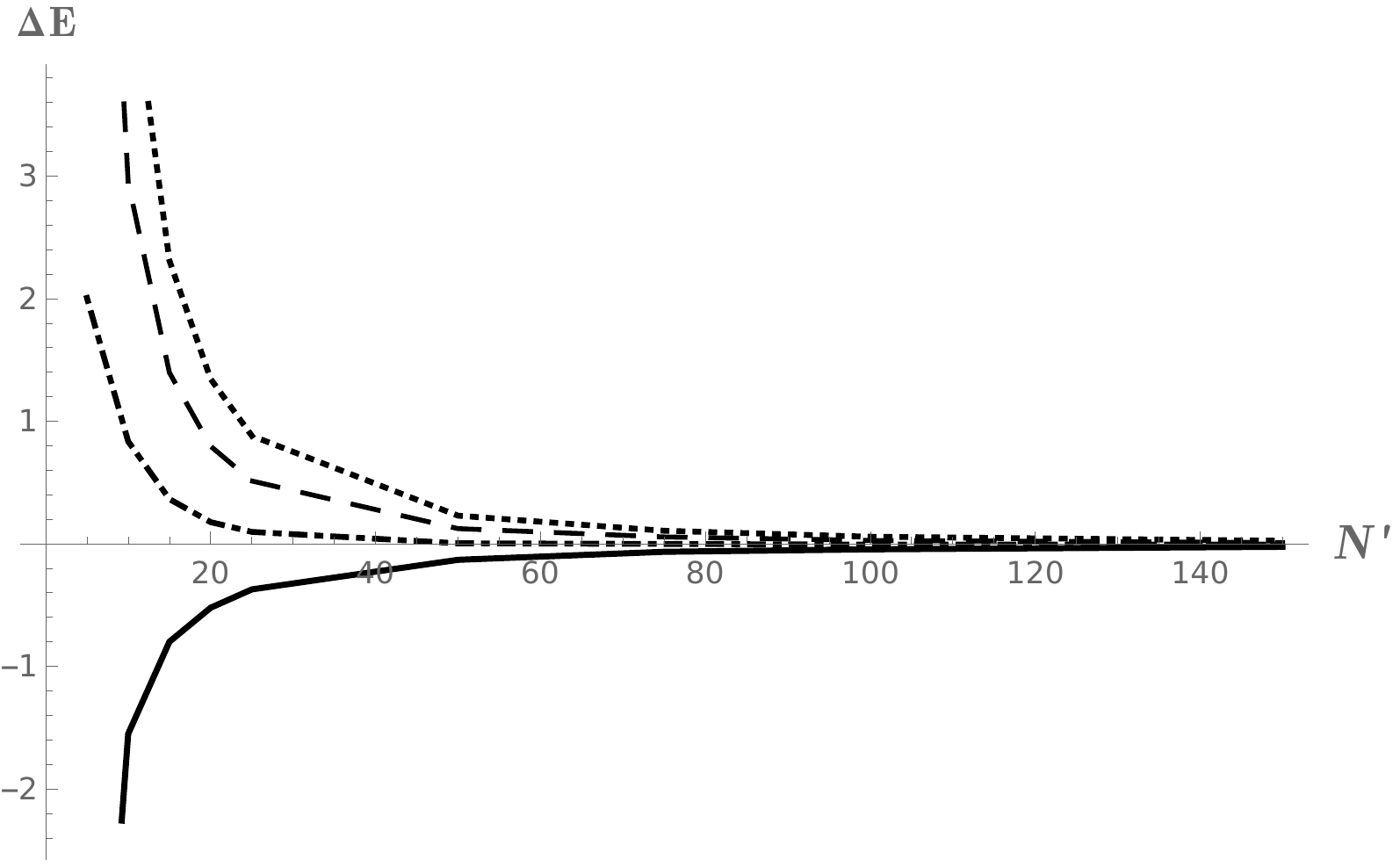}
\caption{Results for the $4s$ electronic energy state depending on the upper limit of summation $\left(N\right)$ in atomic units (a. u.). The dotted, dashed, dot$-$dashed and black lines represents $E\left(h, \nu_{1}\right)$ with values of $\nu_{1}$ obtained from $2^{th}$ $\left(i=2\right)$, $3^{th}$ $\left(i=3\right)$, $4^{th}$ $\left(i=4\right)$, $5^{th}$ $\left(i=5\right)$ iterations, respectively. $\Delta E=\vert E_{exact}\vert-\vert E\left(h,\nu_{1}\right) \vert$. $N'=N/100$. $E_{N}\left(h,\nu_{1}\right) \equiv E_{4s}\left[h,\mu\left(N,l,\nu_{1}\right) \right]$. The results are multiplied by $10^6$.}
 \label{fig:E4s}
\end{figure}

The time$-$independent one$-$dimensional Schr{\"o}dinger equation given in the previous section is solved as a sample. The method of solution has been incorporated into a computer program written in the $Mathematica$ programming language. Variationally optimum values for the grid$-$size $h$ are used in calculations. The optimization procedure is implemented in order to test stability of the suggested method.\\
Results for the $4s$ electronic energy state of hydrogen atom depending on the upper limit of summation $\left(N\right)$ are given in the Figure \ref{fig:E4s}. This figure demonstrates how the correct energy eigenvalues to be found by using the procedure presented in the Section \ref{mu}. It can be seen from this Figure that, the first four iterations convergence from above while the fifth convergences from below. This convergence property exposes at which iteration the algorithm should decide to stop. The algorithm of the procedure with an application for ground state energy calculation is given in the Figure \ref{fig:chart}.\\
The energy eigenvalues of hydrogen atom, up to a principal quantum number $n$, $n=10$, are investigated. The results are presented in the Tables \ref{tab:h}, \ref{tab:eh}, \ref{tab:eih}. Variationally optimum values for the grid$-$sizes depending on the upper limit of summation $N$ are given in the Table \ref{tab:h}. They are obtained by minimizing the energy eigenvalues via the Powell's optimization procedure \cite{70_Powell_1964}. In the Tables \ref{tab:eh}, \ref{tab:eih}, the results of computation for ground and excited states are presented. The matrix form approach is used to solve the Eq. (\ref{eq4}) for Coulomb and screened Coulomb potentials, respectively. They are compared with the results given in \cite{5_Uria_1996} obtained via Richardson's extrapolation and with the results obtained via band matrix technique that recently have been published in \cite{69_Purevkhuu_2021}. The most accurate results that the standard Numerov method with error term is of order $\mathcal{O}\left(h^4\right)$ can provide, were presented in \cite{69_Purevkhuu_2021}. The eigenvalue problem for large  matrices must be overcome for such work. The modified subroutines from the LINPACK library: \textbf{dgbsl.f}, \textbf{dgbfa.f} were accordingly, performed in \cite{69_Purevkhuu_2021}. Fairly large number of grid$-$points (i.e., $\approx 25000$) were used. It was reported that the Numerov's method for $s$ state with error term only is of order $\mathcal{O}\left(h^2\right)$ due to singularity of Coulomb potential. Note that, for the $p$ states is $\mathcal{O}\left(h^3\right)$ and for the $d$ states is $\mathcal{O}\left(h^4\right)$.\\
The results given in the Tables \ref{tab:eh}, \ref{tab:eih} are presented for $s$ states, accordingly. Although less number of grid$-$points are used, more accurate than the ones presented in both Table \ref{tab:eh} and \cite{69_Purevkhuu_2021} are given the Table \ref{tab:eih}. In this table, instead of using large number of grid$-$points as in \cite{69_Purevkhuu_2021}, the range of values of $\nu$ are reduced. The following conclusions are achieved consequently:
\begin{itemize}
\item{The method suggested in this study accelerating the convergence of numerical power series related with the Numerov method.}
\item{Difficulty of obtaining the eigenvalues for large matrices is eliminated.}
\item{The standart Numerov method is limited to error term is of order $\mathcal{O}\left(h^4\right)$. Table \ref{tab:eih} show that, the technique developed here, improves the accuracy for any energy state. It permits to find the energy eigenvalues with more then $\mathcal{O}\left(h^4\right)$ even for potential with a singularity.}
\item{Extension of the Numerov method to error term is of arbitrary order is problematic because $B^{-1}A$ in the Eq. (\ref{eq19}) is not sparse and not symmetric \cite{33_Kuenzer_2019}. Such necessity may also be eliminated. Instead, the range of values of $\nu$ may be further constrained.}
\end{itemize}

Solution of Hartree$-$Fock equations for atoms and molecules provides essential input information in investigating the internal structures and reaction dynamics of complex systems in fields like condensed matter physics, quantum chemistry, and plasma physics. \cite{71_Froese_1963, 72_Fischer_1978, 73_Jiao_2019}. The Hartree$-$Fock equations are solved by self$-$consistent field method. It is an iterative solution obtained by rewriting each Hartree$-$Fock equation in the form of the Schr{\"o}dinger equation with the non$-$local potential. In addition to field of studies referred in the Section \ref{introduction}, the convergence acceleration method for numerical power series suggested in this work is also useful to improve the numerical solutions for Hartree$-$Fock equations.\\
The computational aspect of the formulae given in the present work for numerical calculation of the two$-$ and three$-$dimensional Schr{\"o}dinger equations  will be the subject of the next work.
\section*{Acknowledgement}
In this study, the authors were supported by the Scientific Research Coordination Unit of Pamukkale University under the project number 2020BSP011. One of the author Z. G. is an undergraduate student works under the supervision of A. B. She would like to express her gratitude to the Pamukkale University, Department of Physics for the support during her minor program in physics.

\end{document}